\documentclass[11pt,twoside]{article}
\usepackage{./asp2014}

\aspSuppressVolSlug
\resetcounters

\def\msun{M$_{\odot}$}
\def\inv{$^{-1}$}

\usepackage[usenames,dvipsnames]{xcolor}
\definecolor{blue}{HTML}{4169E1}

\begin{document}

\title{Offset Active Galactic Nuclei}
\author{Blecha, Laura,$^1$, Brisken, Walter$^2$
Burke-Spolaor, Sarah,$^{3,4}$
Civano, Francesca,$^{5}$
Comerford, Julia,$^6$
Darling, Jeremy,$^6$, Lazio, T.~Joseph~W.$^7$, and Maccarone, Thomas J., $^8$
\affil{$^1$University of Florida, Gainesville, FL, USA; \email{lblecha@ufl.edu}}
\affil{$^2$Long Baseline Observatory, Socorro, NM, USA;\email{wbrisken@lbo.us}}
\affil{$^3$Department of Physics and Astronomy, West Virginia University, P.O. Box 6315, Morgantown, WV 26506, USA}
\affil{$^4$Center for Gravitational Waves and Cosmology, West Virginia University, Chestnut Ridge Research Building, Morgantown, WV 26505, USA; \email{sarah.spolaor@mail.wvu.edu}}
\affil{$^5$ Harvard Smithsonian Center for Astrophysics, Cambridge, MA, USA; \email{fcivano@cfa.harvard.edu}}
\affil{$^6$University of Colorado, Boulder, CO, USA; \email{julie.comerford@colorado.edu,jeremy.darling@colorado.edu}}
\affil{$^7$Jet Propulsion Laboratory, California Institute of Technology, Pasadena, CA USA; \email{Joseph.Lazio@jpl.nasa.gov}}
\affil{$^8$Texas Tech University, Lubbock, TX, USA;\email{thomas.maccarone@ttu.edu}}
}

%\affil{$^3$Institution Name, Institution City, State/Province, Country; \email{AuthorEmail@email.edu}}}

% This section is for ADS Processing.  There must be one line per author.
\paperauthor{Laura Blecha}{lblecha@ufl.edu}{}{University of Florida}{Physics}{Gainesville}{FL}{32611}{USA}
\paperauthor{Walter Brisken}{wbrisken@lbo.us}{}{Long Baseline Observatory}{}{Socorro}{NM}{}{USA}
\paperauthor{Sarah Burke-Spolaor}{sarah.spolaor@mail.wvu.edu}{}{West Virginia University}{Physics and Astronomy}{Morgantown}{WV}{26505}{USA}
\paperauthor{Francesca Civano}{fcivano@cfa.harvard.edu}{}{Harvard Smithsonian Center for Astrophysics}{}{Cambridge}{MA}{}{USA}
\paperauthor{Julia Comerford}{julie.comerford@colorado.edu}{}{University of Colorado}{}{Boulder}{CO}{}{USA}
\paperauthor{Jeremy Darling}{jeremy.darling@colorado.edu}{}{University of Colorado}{}{Boulder}{CO}{}{USA}
\paperauthor{T.~Joseph~W. Lazio}{Joseph.Lazio@jpl.nasa.gov}{}{Jet Propulsion Laboratory, California Institute of Technology}{}{Pasadena}{CA}{}{USA}
\paperauthor{Thomas J. Maccarone}{thomas.maccarone@ttu.edu}{}{Texas Tech University}{}{Lubbock}{TX}{}{USA}

%\paperauthor{Sample~Author3}{Author3Email@email.edu}{ORCID_Or_Blank}{Author3 Institution}{Author3 Department}{City}{State/Province}{Postal Code}{Country}

\begin{abstract}
Gravitational-wave (GW) and gravitational slingshot recoil kicks, which are natural products of SMBH evolution in merging galaxies, can produce active galactic ``nuclei'' that are offset from the centers of their host galaxies. Detections of offset AGN would provide key constraints on SMBH binary mass and spin evolution and on GW event rates. Although numerous offset AGN candidates have been identified, none have been definitively confirmed. The ngVLA offers unparalleled capabilities to identify and confirm candidate offset AGN from sub-parsec to kiloparsec scales, opening a new avenue for multi-messenger studies in the dawn of low-frequency GW astronomy.
\end{abstract}

\section{Offset AGN as Signposts for Supermassive Black Hole Mergers}

High-angular-resolution, high-sensitivity radio observations with the Next Generation Very Large Array (ngVLA) can uniquely probe a poorly understood aspect of supermassive black hole (SMBH) evolution: how often are SMBHs offset from the nuclei of their host galaxies? Our nomenclature for accreting SMBHs---namely, active galactic nuclei, or AGN---reflects their nominal  location, but SMBHs are not always centrally located. In ongoing galaxy mergers, for example, offset AGN are produced when one of the two SMBHs is active; many such objects have been found (e.g., \citealt{barrow16}). Here we focus instead on other sources of offset SMBHs: gravitational-wave (GW) recoil and gravitational slingshot kicks, which can displace SMBHs from nuclei or even eject them from galaxies entirely. Such systems are more than a mere curiosity. Confirmed offset SMBHs can constrain the evolution of binary and merging SMBHs, including mass and spin evolution in the final stages of the merger. Ejected and offset SMBHs also affect BH-galaxy co-evolution and introduce scatter into BH-bulge correlations \citep{volont07,blecha11,sijack11}.

\subsection{Recoiling Black Holes}
\begin{figure*}
\centering
\includegraphics[width=0.22\textwidth]{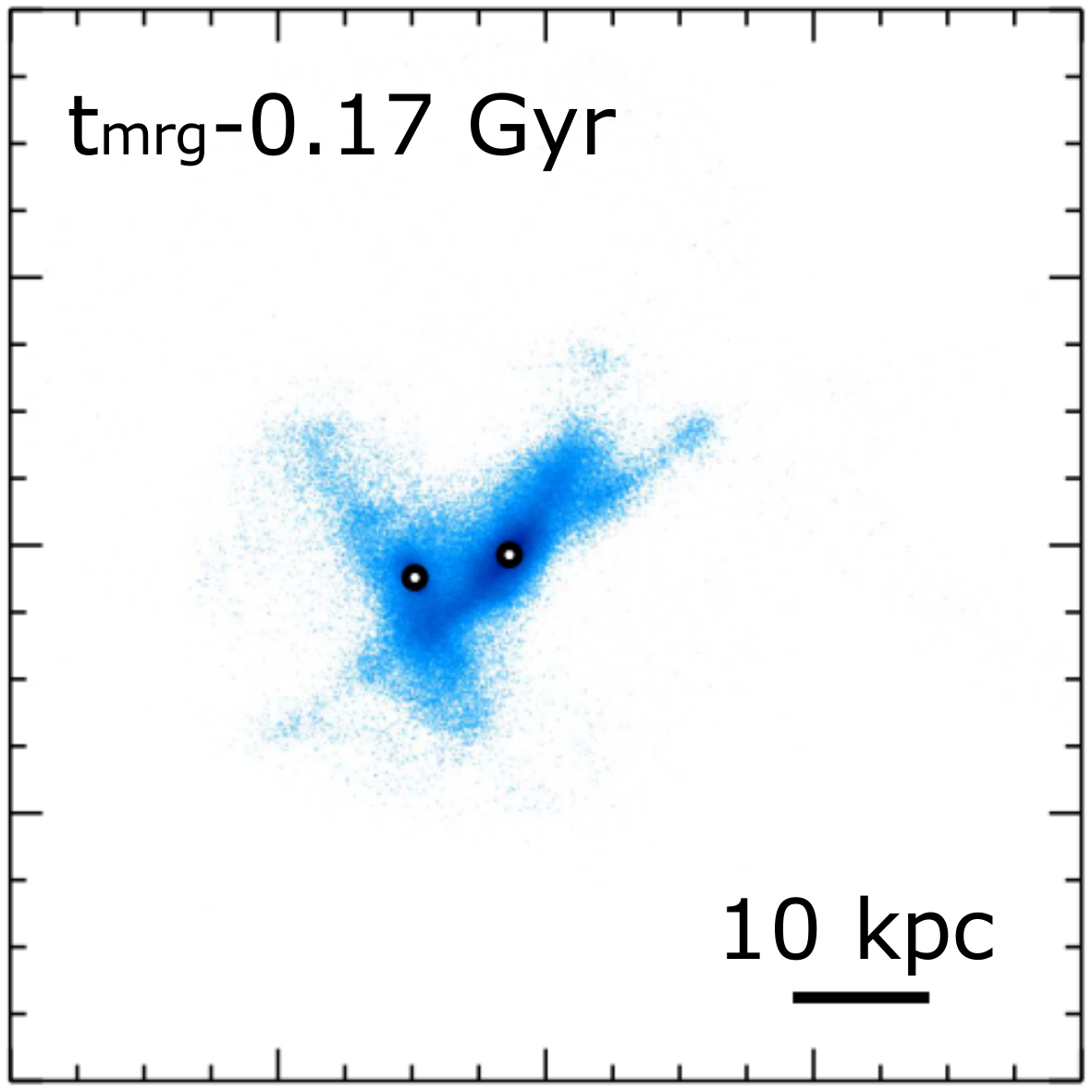}
\includegraphics[width=0.22\textwidth]{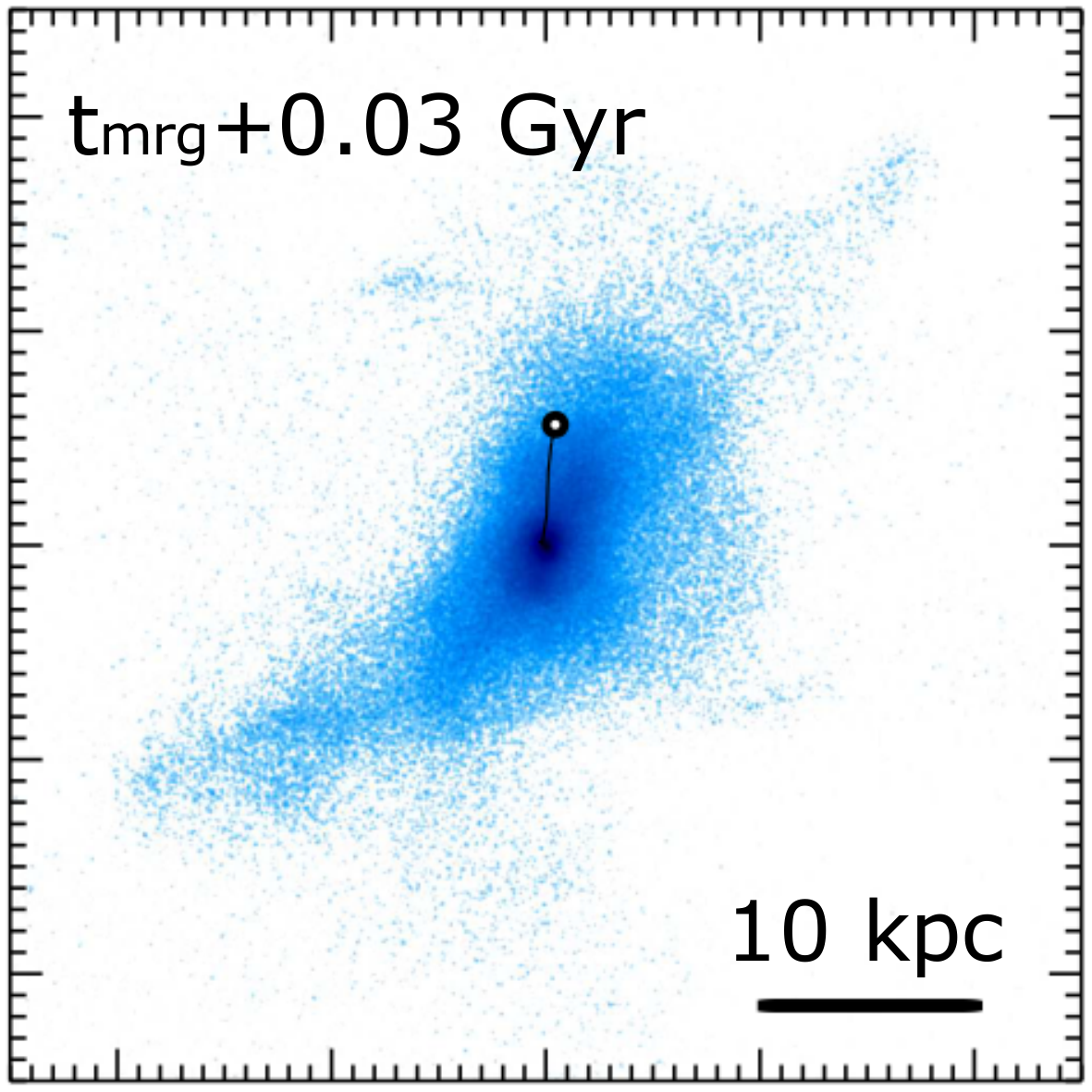}
\includegraphics[width=0.22\textwidth]{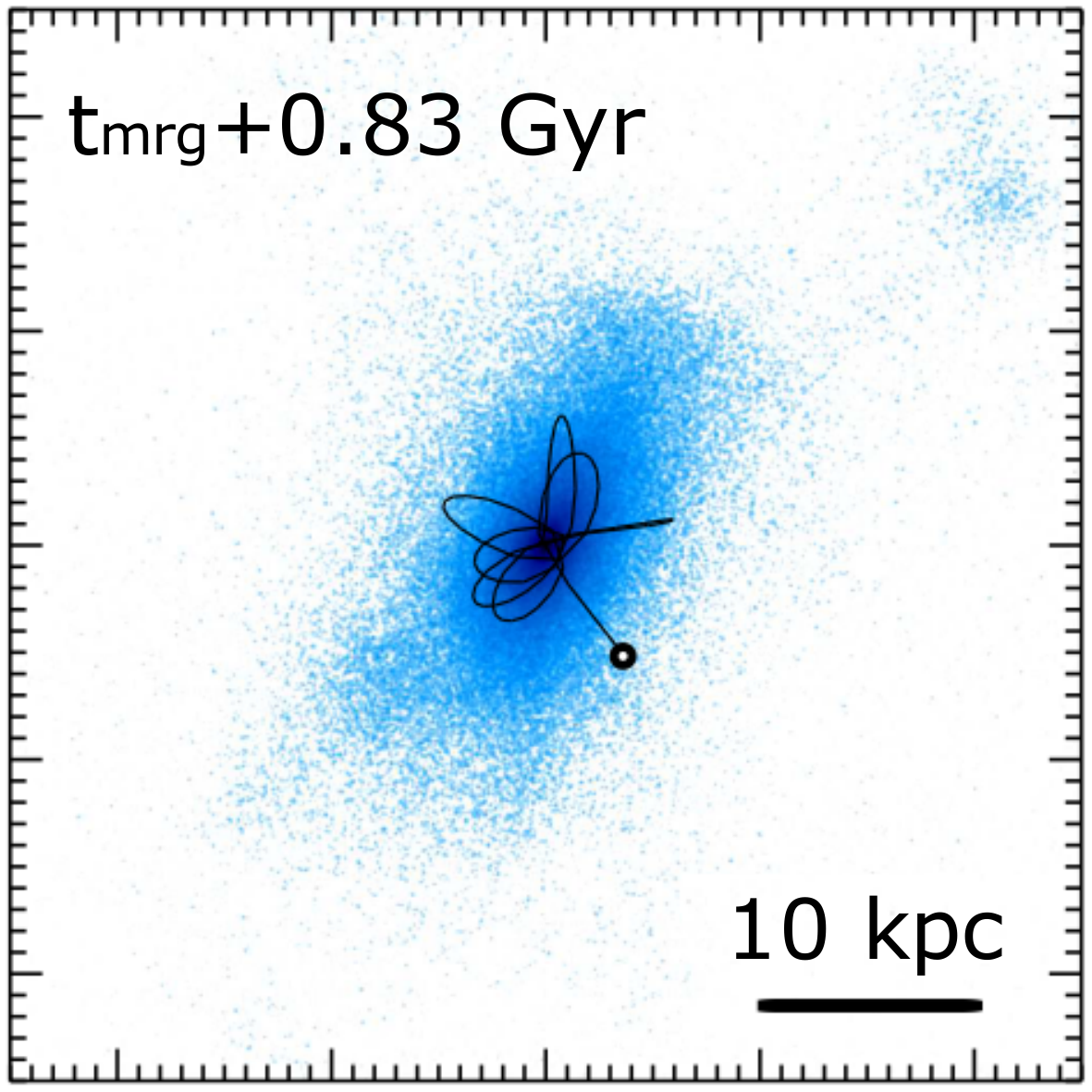}
\includegraphics[width=0.22\textwidth]{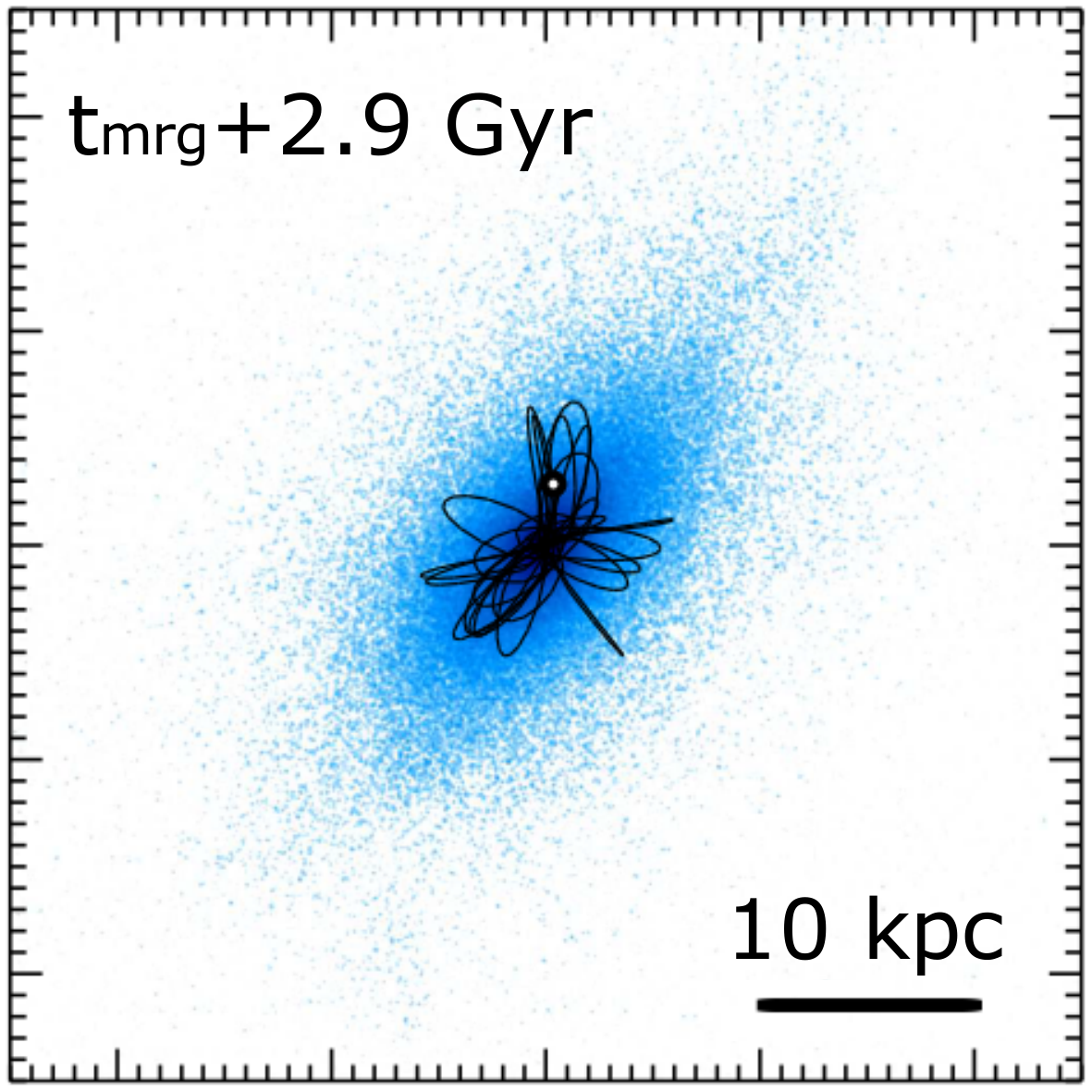}
\caption{Illustrative example of merging galaxies and subsequent recoiling SMBH trajectory at four time snapshots, from a {\footnotesize GADGET} hydrodynamics simulation of an isolated galaxy merger from \citet{blecha11}. The blue color scale shows the stellar density; black dots denote the SMBH position. The left panel shows the two galaxy nuclei shortly before coalescence; subsequent panels show the (empty) merged nucleus and the trajectory of the offset recoiling SMBH after being kicked at 80\% of the central escape speed. \label{fig:recoil_traj}}
\end{figure*}

Asymmetric gravitational wave (GW) emission during a BH merger can impart a kick of up to 5000~km~s\inv\ to the BH merger remnant \citep{campan07b, lousto10}, which can eject SMBHs from even galaxy clusters. Such extreme recoil kicks should be exceedingly rare, but kicks of even a few hundred km s\inv\ can cause SMBHs to wander for more than a Gyr and can produce detectable offsets (e.g., \citealt{guamer08}, \citealt{bleloe08}; Figure \ref{fig:recoil_traj}). GW recoil velocities depend sensitively on the BH spins just before merger---if the BHs are non-spinning or have perfectly aligned spins, the maximum possible kick is only 175~km~s\inv. Few constraints on SMBH spins currently exist, but detections of offset, rapidly-recoiling SMBHs would provide strong evidence for misaligned, spinning progenitor SMBHs.

Alternately, if binary SMBH inspiral timescales are longer than the typical time between galaxy mergers, a subsequent galaxy merger may introduce a third BH into an existing binary system. If a close triple system forms, the three-body encounter will often eject the lightest SMBH, producing a gravitational slingshot kick and possibly driving the remnant binary SMBH to rapid merger \citep[e.g.,][]{hofloe07,bonett16,bonett18}. A key distinction between GW and slingshot recoil is that the latter leaves more than one BH remaining in the system after the kick.

\subsection{Observable Signatures of Offset AGN}
In either case, if the recoiling SMBH is actively accreting at the time of the kick, it will carry along its accretion disk, broad emission line (BL) region, and radio-emitting core (for most recoils, everything within $\sim 10^{4}$--$10^{5}$ gravitational radii will typically remain bound to the SMBH). The recoiling SMBH could then be observed as an ``offset AGN'' for up to tens of Myr \citep[e.g.,][]{madqua04,loeb07,blecha11}. In cases where the SMBH is not ejected from the galaxy entirely, larger offsets are still favored, because it spends most time near the apocenter of its orbit. Depending on the recoil trajectory and the type of observation, spatial and/or velocity offsets could be observed. 

Unambiguous confirmation of recoiling AGN has proven tricky with existing observations due to a number of factors, including the large samples needed for a discovery, possible confusion with an inspiraling pre-merger AGN, and the limited resolution and sensitivity of infrared (IR), optical, and ultraviolet (UV) telescopes. The ngVLA will be able to resolve spatial offsets at sub-pc scales, such that the sensitivity to GW recoiling AGN will be limited primarily by the accuracy of relative astrometry with optical or IR imaging of the host galaxy centroid. At least when host nuclei are comprised mainly of older stellar populations, astrometric centroiding in the near-IR can be quite accurate. \citet{condon17} conducted a search for offset AGN with the VLBA, using 2MASS data to centroid the host bulge to within $\sim 100$ mas. They found an offset AGN traveling away from a brightest cluster galaxy (BCG) core at $\ga 2000$ km s\inv, but they were able to identify a secondary compact galaxy associated with the offset AGN; this and surrounding tidal debris strongly indicate tidal stripping of a satellite galaxy by the BCG rather than a recoiling AGN. 

In contrast, a slingshot recoil, in which a close binary SMBH (or single merged SMBH) remains behind as a dual or single radio source, could be resolved down to the limit of relative ngVLA astrometry ($\la$ 1\% of the beam FWHM for moderately bright sources). Moreover, the ngVLA offers the entirely new possibility of measuring the proper motions of rapidly-recoiling AGN in nearby systems.  A ngVLA detection of a displaced AGN with a proper motion measurement would constitute the gold standard for a confirmed recoiling AGN.

\begin{figure}
\centering
\includegraphics[width=0.28\textwidth]{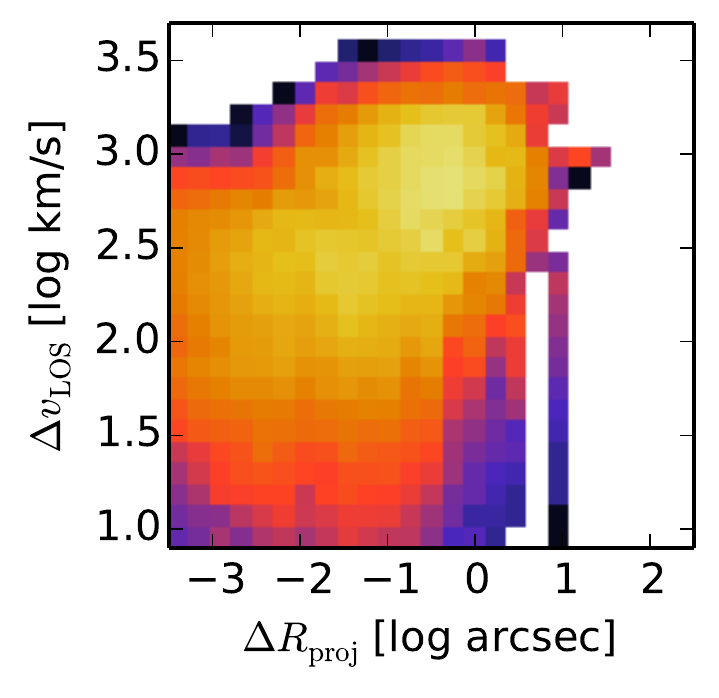}
\includegraphics[width=0.28\textwidth]{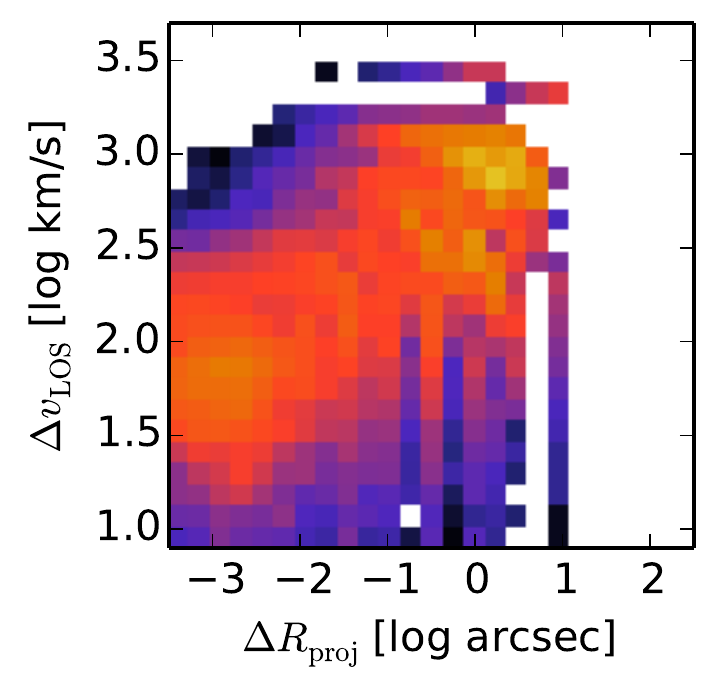}
\includegraphics[width=0.39\textwidth,trim = 6 0 0 6]{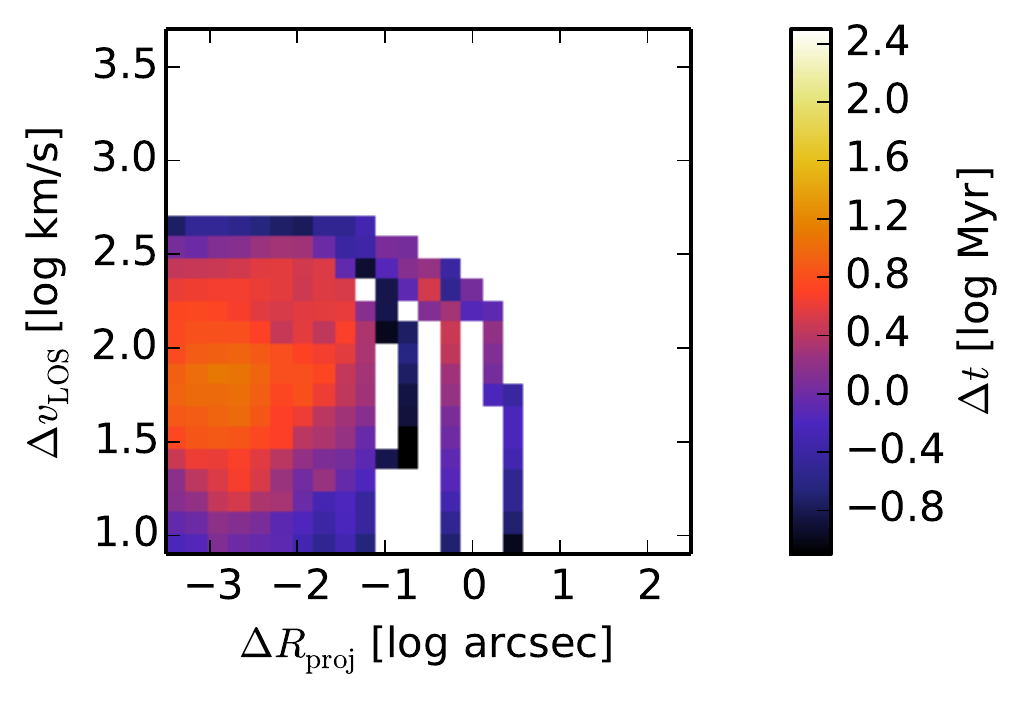}
\caption{Time-weighted distributions of recoiling AGN offsets across cosmic time, based the Illustris cosmological simulations and the recoil model of \citet{blecha16}. Only GW recoil is considered (not slingshot recoil), and only AGN that would be optically detected at the depth of HST-COSMOS are included. 
{\em Left:} Pre-merger spins are randomly oriented.
{\em Middle:}  A ``hybrid'' spin model is assumed in which BH spins are aligned only in gas-rich mergers. 
{\em Right:} Pre-merger BH spins are aligned to within $5^{\circ}$. The ngVLA will be sensitive to recoiling AGN with both small and large spatial offsets, and for highly aligned BH spins, GW recoils may be detectable only with long-baseline radio imaging. ngVLA detections of spatial offsets coupled with line-of-sight velocity offsets from optical spectra would provide strong evidence for recoiling BHs. (From \citealt{blecha16}.)\label{fig:offsets}}
\end{figure}

The BL region remains gravitationally bound to the recoiling SMBH, while the narrow emission line region (on $\la$ kpc scales) is left behind in the galaxy. Consequently, one observable signature of a recoiling SMBH would be a velocity offset between the broad and narrow lines. The relative rarity of large BL offsets in SDSS quasars \citep{eracle12} is in tension with the predicted number of observable recoiling AGN, {\em if} most progenitor SMBHs have randomly oriented spins \citep{blecha16}. This suggests that spin alignment occurs in some merging SMBH binaries. Assuming that gas-rich galaxy mergers form massive circumbinary disks that torque BH spins into alignment, but that gas-poor mergers have misaligned spins, the cosmological models of \citet{blecha16} predict that hundreds of spatially offset AGN may be detected at the sensitivity and resolution of large-area surveys with, e.g., LSST, WFIRST, or Euclid \citep[see also][]{volmad08}. Figure \ref{fig:offsets} suggests that a ngVLA survey could detect a much higher space density of recoiling AGN with smaller spatial offsets. Moreover, these predictions for optical/near-IR surveys do not account for possible nuclear obscuration, which is prevalent in merger-triggered AGN hosts \citep[e.g.,][]{veille09,kocevs15,ricci17,blecha18}. While nuclear obscuration in the remnant galaxy will not affect recoiling AGN with large spatial offsets, the ngVLA will be uniquely sensitive to offset AGN on $<$ pc-to-kpc scales that are still embedded in obscuring nuclear gas and dust.

\section{Current Searches for Offset AGN and Limitations}

\subsection{Spectroscopic and kpc-scale Spatial Offsets}
A growing number of candidate recoiling AGN have been found over the last decade. Some have been identified spectroscopically, via offset BLs separated by $\ga 1000$~km~s\inv\ from the narrow emission lines \citep[e.g.,][]{komoss08, robins10}. \citet{eracle12} have identified a sample of 88 SDSS quasars with large BL offsets; some of these could be interpreted as recoiling AGN candidates, though offset BLs can also arise from binary SMBHs, unusual double-peaked emitters, or gas outflows \citep[e.g.,][]{decarl14,runnoe17}.

Other recoil candidates have been identified via spatial offsets on kpc scales \citep{jonker10, koss14, markak15,kalfou17}. If an offset AGN is well-separated from a galaxy and is not embedded in a secondary galaxy of its own, it cannot be explained by dual BH scenarios. However, if the host galaxy of an inspiraling, offset AGN has been significantly tidally stripped by a larger galaxy (as with some ultra-compact dwarfs; e.g., \citealt{seth14}, or in the cores of galaxy clusters; e.g., \citealt{condon17}), the offset host galaxy and tidal streams may be difficult to detect beyond the local Universe. For nearby offset AGN, high-resolution optical or near-IR imaging can place stringent constraints on the presence of an associated stellar nucleus component that is separate from the primary (central) galaxy \citep[e.g.,][]{koss14}. Note that the presence or absence of disturbed morphology in a galaxy does not itself distinguish between a recoiling SMBH and an inspiraling dual SMBH, as tidal features are typically present for hundreds of Myr before and after the SMBH merger.

It is also possible for unusual long-lived Type~IIn supernovae to mimic offset AGN signatures, however; SDSS 1133 is a striking example of such a candidate \citep{koss14}. Further monitoring and high-resolution follow-up should be able to distinguish between a recoiling AGN and a Type IIn SN in the near future. More generally, time domain surveys such as Pan-STARRS and LSST can be used to identify offset point sources and distinguish those that are stochastically variable from transient SN flares \citep{kumar15}.

\begin{figure}[tb]
\centering
\includegraphics[width=0.462\textwidth]{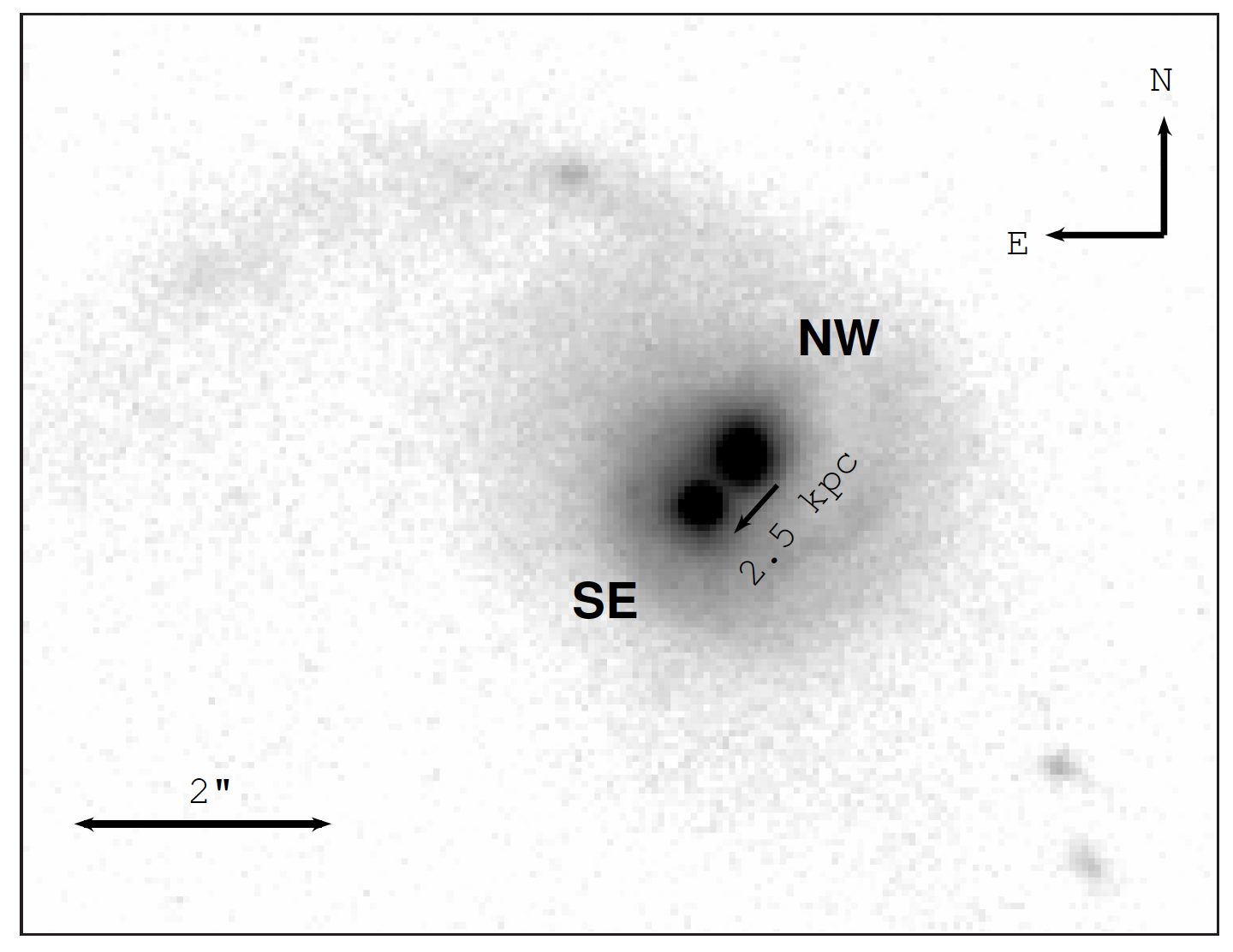}
\includegraphics[width=0.49\textwidth,trim={0cm 0cm 0cm 5cm},clip]{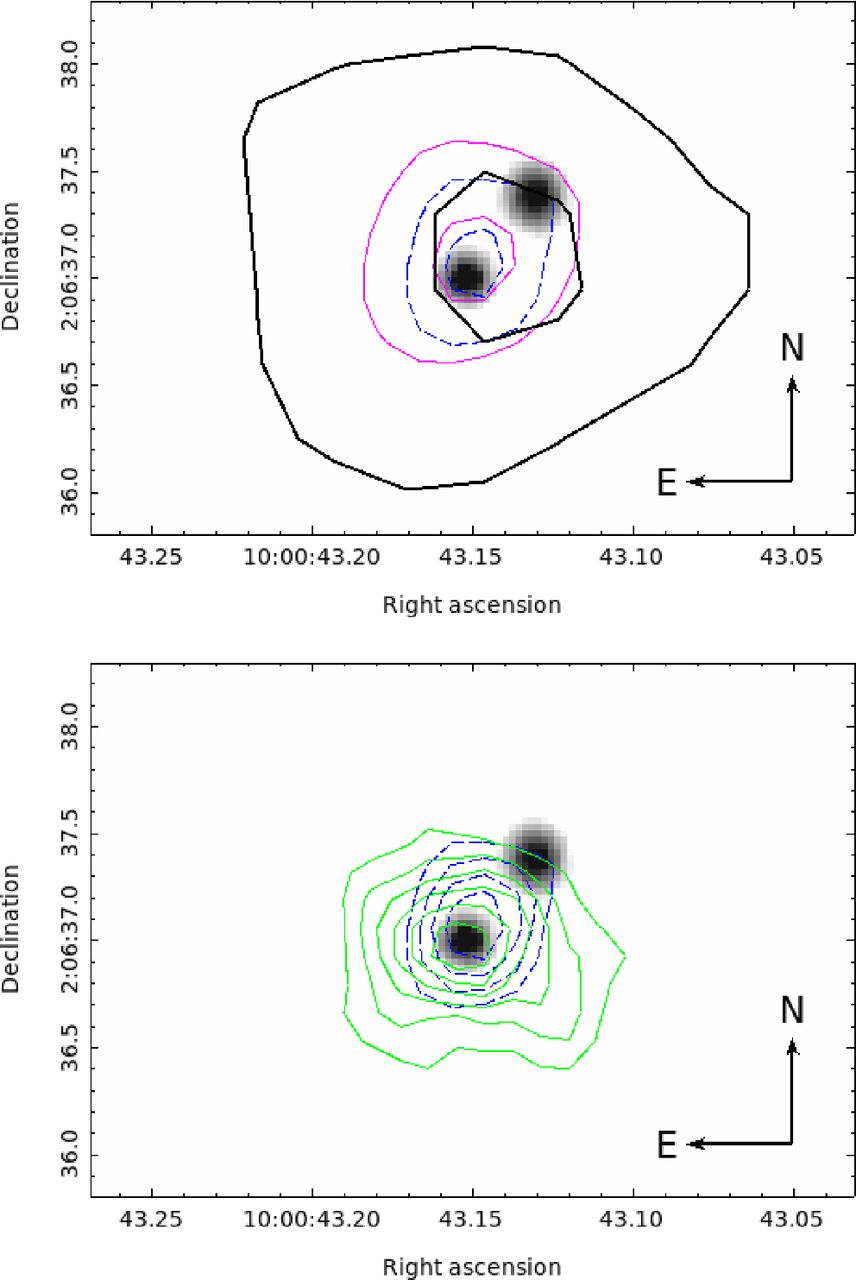}
\includegraphics[width=0.7\textwidth]{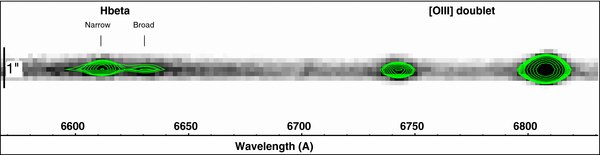}
\caption{Observations of recoiling AGN candidate CID-42. {\em Top left:} The F814W {\em HST} image of this source reveals two optical nuclei separated by 2.5 kpc and a prominent tidal tail. The NW (upper right) nucleus is marginally resolved; the SE (lower left) nucleus is a point source. (From \citealt{civano10}.) {\em Top right:} Combined image of the nucleus of CID-42 including HST optical (gray), Chandra X-ray (green), and VLA 3 GHz radio (blue). The X-ray and radio emission are consistent with a point source originating entirely from the SE source, which is the putative recoiling AGN. (From \citealt{civano12} and \citealt{novak15}.) {\em Bottom:} Magellan/IMACS optical slit spectrum of CID-42, showing an offset of 1300 km s\inv\ between the broad and narrow lines. (From \citealt{civano10}.) \label{fig:candidates}}
\end{figure}

Arguably the most promising candidates are those with both spatial and velocity offset signatures \citep{civano10, civano12, chiabe17}; CID-42 is one such example (Figure \ref{fig:candidates}). Simulations indicate that this system can be well-modeled as a rapidly recoiling BH with a plausible offset AGN lifetime \citep{blecha13a}. These combined lines of evidence present a compelling case for a rapidly recoiling BH in CID-42, but the data could also be explained as an inspiraling, kpc-scale BH pair in which one BH is quiescent or intrinsically faint. 

In fact, none of the recoil candidates discovered to date has been confirmed. If the recoil candidate is superimposed on the host galaxy, the dual BH scenario is difficult to exclude, but high-quality resolved stellar or gas kinematics (with, e.g., ALMA, JWST or optical integral field spectroscopy) could distinguish between a pre- and post-merger nucleus. High-sensitivity ngVLA observations will provide strong constraints on AGN offsets and on the possible presence of a secondary, faint AGN in the host nucleus.

\subsection{Parsec-scale Spatial Offsets}
If a recoiling BH is not ejected entirely from its host galaxy, it will eventually return to the nucleus, possibly several Gyr after the recoil event, where it may undergo long-lived, small-amplitude oscillations about the galaxy center \citep[e.g.,][]{guamer08, bleloe08}. \citet{batche10} and \citet{lena14} found significant optical photometric offsets between the AGN and galaxy nucleus on scales $< 10$ pc in six core elliptical galaxies (out of 14 observed, including M87), suggesting that such small-scale offsets are quite common in core ellipticals. However, in four of these galaxies, the offset is approximately aligned with the axis of coincident radio emission, which cannot be readily explained by a physical mechanism. And in fact, the measured offset in M87 has recently been attributed to a transient jet outburst; more recent data show no evidence of a spatial offset \citep{loppri18}. In order to determine the true prevalence of small-scale AGN offsets, high-sensitivity observations ngVLA are needed to probe low-luminosity and radio-quiet AGN in galaxies with diverse morphologies.

\section{Unique ngVLA Capabilities}
The high resolution, high sensitivity, and broad frequency coverage of the ngVLA will enable a new class of observations to identify and follow up candidate recoiling AGN. In particular, the increase in resolution relative to the VLA along with increased sensitivity will allow more rapid detection of small-scale spatial offsets and small-scale extended features in a large number of objects, which will be critical for both systematic searches and for follow up of offset AGN candidates. The increase in sensitivity over the VLA and VLBA will allow the first large systematic radio search for offset AGN, and for known candidates, the ngVLA will place stringent limits on low-surface-brightness extended features and secondary sources. The broadband spectral coverage of the ngVLA combined with these capabilities will provide crucial data for confirming or eliminating candidate recoiling AGN in cases where their true nature cannot be discerned via other means. We describe several promising possible studies here:\\

\noindent {\bf \em Survey for spatially-offset AGN}

The ngVLA will have the capability to carry out a systematic survey of thousands of galaxies with high sensitivity, which will enable a search for compact radio cores offset from their host nuclei on sub-pc to kpc scales. Such a survey would be well-aligned with the goals of a complementary search for $\sim$ pc-scale and sub-pc-scale binary SMBHs (see chapter on SMBH binaries in this volume; Burke-Spolaor et.~al.). High dynamic range observations with the ngVLA would allow either detections of dual compact radio cores with unequal luminosities or strong limits on the absence of a second radio core; the latter would indicate a possible GW recoiling SMBH. Furthermore, unlike optical instruments, the ngVLA will be able to detect AGN with small-scale ($<$ pc - kpc) offsets that are embedded in heavily obscured nuclei, which are common in merger remnants. Because the accretion rate will decrease monotonically with time for an isolated, recoiling AGN, high sensitivity observations with the ngVLA can also detect offset AGN that have traveled to separations $\ga 1$--10~kpc from the host nucleus. 

Based on the models of \citet[][Figure \ref{fig:offsets}]{blecha16}, we estimate that spatially offset, GW-recoiling AGN detectable with the ngVLA (with a SNR of 10 and spatial resolution of 10 mas) have a density of $> 0.1$ deg$^{-2}$ out to $z\sim 0.5$. This of course assumes that the host centroid can be measured with sufficient accuracy, which would be difficult for small offsets at these redshifts. In the nearby universe ($z \la 0.05$), we estimate that up to $\sim$ 60-70 spatially offset GW-recoiling AGN could be detected with the ngVLA under the same assumptions. This also assumes that SMBH binary spins are not always highly aligned prior to merger. 

If spins {\em are} always efficiently aligned to within a few degrees, GW recoil velocities are much lower ($\la 300$ km s\inv), and spatial offsets $\ga 0.05"$ are very rare (Figure \ref{fig:offsets}). Crucially, this means that long-baseline ngVLA observations are about the {\em only} means by which spatially offset, GW-recoiling AGN could be detected. If spins are always aligned to within $5^{\circ}$, then using the same assumptions as above we expect 1--2 detectable spatially offset AGN within $z<0.05$, and $>0.01$ deg$^{-2}$ out to $z\sim0.5$. The spatial offsets of any GW-recoiling AGN identified with the ngVLA would therefore constrain the pre-merger spins of SMBH binaries.

Time variability and spectral information can be used to determine whether an offset radio source is a true offset radio AGN core or simply a knot in a radio jet. Specifically, spectral index measurements can distinguish between a flat-spectrum core (indicating ongoing AGN activity) and a steep-spectrum jet (with an older electron population). The wide frequency coverage and high resolution of the ngVLA is ideal for this kind of measurement, and its broadband capabilities allow spectral indices to be measured within a single band.

Galaxies with a single apparent stellar nucleus that also have disturbed morphology will be promising targets to search for offset or dual radio cores. Core elliptical galaxies with shallow central density profiles are also promising targets, as recoil oscillations about the nucleus should be longer lived in such systems; however, multi-wavelength astrometry of the host nucleus is more difficult in these cases. Low-mass galaxies are of great interest as well, because of their potential to constrain the population of isolated and merging low-mass SMBHs that are most relevant for LISA. Although the galaxy merger rate is lower for low-mass galaxies, their lower escape speeds greatly increase the probability that a recoil kick will produce a measurable spatial offset. \\

\noindent {\bf \em Follow-up of known candidates}

The ngVLA will provide an excellent means of following up known recoil candidates. In many cases, high dynamic range, high sensitivity radio observations will be crucial for distinguishing between a kpc-scale, inspiraling dual AGN (in which one AGN is intrinsically faint) and a single, offset recoiling AGN (cf.\ CID-42, Figure~\ref{fig:candidates}). Even when the offset source is well-separated from the galaxy, detection of a compact radio core coincident with an optical, IR, or X-ray point source would provide strong support for the recoil scenario, and would be crucial for confirming offsets that are marginally resolved at other wavelengths.

LSST, WFIRST, and Euclid have the potential to identify a large population of spatially offset AGN \citep{blecha16}. LSST will also be able to distinguish stochastically-variable offset AGN from transient offset sources \citep[i.e., SNe; cf.][]{kumar15}. ngVLA variability studies will also be important for confirmation of recoiling AGN candidates identified in single- or multi-epoch data, and they can be used to rule out apparent small-scale spatial offsets produced by transient jet phenomena \citep[cf.][]{loppri18}. 

In addition to the sample of known quasars with large BL offsets \citep{eracle12}, more such candidates will be identifiable in newer spectroscopic surveys (e.g., SDSS eBOSS and DESI). ngVLA observations of these sources will be uniquely capable of resolving pc-scale binary SMBHs or detecting offset SMBHs on pc to kpc scales. \\

\noindent {\bf \em Proper motion measurements }

Long baselines ($\ga$ 1000 km) would make the ngVLA uniquely capable of providing the most direct type of confirmation of a recoiling AGN: proper motion studies of recoil. Relative astrometric precision of $< 1$\% of the beam FWHM would enable proper motions of $\sim 1 \mu$as yr\inv\ to be detected out to $\sim 200$ Mpc over a 5--10~yr time baseline for transverse recoil velocities $\ga 1000$~km~s\inv. We predict that up to $\sim 10$ detectable recoiling AGN with such a high transverse velocity could be found within this volume (if spins are not always efficiently aligned prior to merger---no high-velocity recoils will occur in that case). 

In practice, achieving this precision for single objects (i.e. GW recoiling AGN) requires accounting for other sources of proper motion that can be of similar magnitude, such as secular aberration drift, secular extragalactic parallax, transverse peculiar motion, and jet plasma motion. However, peculiar motion of the host galaxy will generally be much less than 1000 km s\inv\ except in cluster environments, and other sources of secular proper motion can be accounted for in cases where a nearby source is present. Repeated multi-band observations of promising candidates can also be used to correct for intrinsic motion of a jet. Moreover, in nearby galaxies with high star formation rates (tens of \msun\ yr\inv, as may be expected in the final stages of a gas-rich major merger hosting a recoiling SMBH), we expect tens of HII regions to be luminous enough for $\la 10$ $\mu$as astrometric precision, which would greatly constrain the bulk proper motion of the galaxy and reduce uncertainty on the relative proper motion of the AGN.

Slingshot recoils present easier targets for proper motion studies. The single or binary SMBH remaining in the galaxy nucleus would allow precise relative astrometry with the full capabilities of the long-baseline ngVLA, such that high-velocity proper motions out to $\sim 200$ Mpc could be more easily identified. The prevalence of slingshot recoils resulting from three-body SMBH interactions is not known, but theoretical studies suggest that they are not uncommon \citep[][]{kelley17a,kulloe12}. Promising monitoring targets would be galaxies that exhibit morphological and kinematic signatures of recent mergers (such as tidal shells and streams) in which two radio AGN are at least marginally resolved. Because slingshot recoil requires that a second galaxy merger occurs before the SMBH binary from a previous merger has had time to coalesce, likely hosts should also include gas-poor, bulge-dominated galaxies in which SMBH binary inspiral is likely to stall, as well as galaxies in rich group environments.\\

\noindent {\bf \em Searches for LISA electromagnetic counterparts}

A large fraction of merging SMBHs should be in gas-rich
environments where they can be expected to accrete near the Eddington
luminosity \citep[e.g.,][]{bode10, farris14, bowen17}.  Because LISA
will rarely give adequate time to look for massive merging SMBHs
beforehand, an ``after-the-fact'' technique for
identifying the EM counterparts is essential for
finding the true host galaxies and thereby using these events for standard siren measurements of the expansion of the Universe.
Eddington-limited $10^7$ M$_\odot$ SMBHs at $z\approx{1}$ should
have core radio flux densities of $\approx{10}$ $\mu$Jy, meaning that
$\sim100$ of them could be well-measured in 10 hours with the ngVLA.
Because the SMBH merger product will be less massive that the sum of the
two progenitor SMBHs, the accretion disk will temporarily fail
to reach the innermost stable circular orbit (ISCO) of the new system and will therefore suddenly cease to produce a strong jet; accretion will restart on the much longer (many-year) viscous timescale. The amplitude of this variability should be far larger than typically seen from normal AGN activity. Monitoring could then be used to identify the object whose core turns
off completely, starting at high frequencies and gradually
progressing to lower frequencies. This is expected to
happen on timescales of a few weeks at 100 GHz and a few years at 1
GHz.  The other high-resolution radio-based technique for identifying the hosts of
recent SMBH mergers is that, if the pre-merger BH spins were
misaligned with one another, there should be two hot spots rather
than one in the AGN lobe.\\

\noindent {\bf \em Other multi-wavelength and multi-messenger synergy}

ngVLA observations will be a crucial component of multi-wavelength campaigns to identify and confirm candidate recoiling AGN. High-resolution optical or IR imaging of the host stellar light is needed to measure the position of the host centroid and to identify signatures of a recent merger (tidal tails, shells, etc.); this includes ground-based adaptive optics imaging with current instruments and future thirty-meter class telescopes, as well as with HST and JWST. Resolved optical, IR, or sub-mm spectroscopy can reveal the nuclear stellar and gas dynamics and better constrain the dynamical center and merger stage of the galaxy. In addition to radio spectral data, multi-wavelength imaging can also be used rule out the possibility of a knot in a radio jet mimicking an offset radio core. Optical spectra also provide BL velocities. X-ray imaging and spectra, particularly in hard X-ray bands, will complement radio observations of offset AGN and constrain the nuclear column density (and hence the possible presence of a second, obscured AGN). 

Any ngVLA detections of recoiling (or binary) SMBHs in advance of LISA science observations can be used to constrain the LISA event rate. Moreover, even non-detections will constrain pre-merger spin evolution and thus the likelihood that LISA GW waveforms will be precessing. ngVLA and other multi-wavelength detections of offset AGN, combined with GW detections of SMBHs binaries and mergers with PTAs and LISA, have the potential to transform our understanding of SMBH evolution in merging galaxies.\\

The ngVLA studies outlined here offer unique capabilities for identifying and confirming offset AGN that cannot be achieved with other observations alone. These studies will constrain the mass and spin evolution of SMBHs in mergers as well as the GW source population. ngVLA searches for recoiling AGN have strong synergy with imaging and spectroscopic observations from sub-mm to X-ray wavelengths, and they are a natural complement to searches for binary SMBHs with the ngVLA and other instruments. The dynamical evolution of SMBHs in merging galaxies is a pressing open question in the dawn of low-frequency GW astronomy, and ngVLA studies of offset AGN will be a key component of the multi-messenger campaign to unravel this mystery.

\acknowledgements LB acknowledges support by NSF award \#1715413.  JD acknowledges support from the NSF grant AST-1411605.  Part of this research was carried
out at the Jet Propulsion Laboratory, California Institute of Technology, under a contract with the National Aeronautics and Space Administration.
% Keep this text on the same line as the \verb"\acknowledgements" command because it makes things a lot easier.

%\bibliography{refs_ngvla}  % For BibTex

\begin{thebibliography}{}
\expandafter\ifx\csname natexlab\endcsname\relax\def\natexlab#1{#1}\fi
\expandafter\ifx\csname url\endcsname\relax
  \def\url#1{\texttt{#1}}\fi
\expandafter\ifx\csname urlprefix\endcsname\relax\def\urlprefix{URL }\fi
\providecommand{\eprint}[2][]{\url{#2}}

\bibitem[{{Barrows} et~al.(2016){Barrows}, {Comerford}, {Greene}, \&
  {Pooley}}]{barrow16}
{Barrows}, R.~S., {Comerford}, J.~M., {Greene}, J.~E., \& {Pooley}, D. 2016,
  \apj, 829, 37. \eprint{1606.01253}

\bibitem[{{Batcheldor} et~al.(2010){Batcheldor}, {Robinson}, {Axon}, {Perlman},
  \& {Merritt}}]{batche10}
{Batcheldor}, D., et.~al.
%{Robinson}, A., {Axon}, D.~J., {Perlman}, E.~S., \&
%  {Merritt}, D. 
  2010, \apjl, 717, L6. \eprint{1005.2173}

\bibitem[{{Blecha} et~al.(2013){Blecha}, {Civano}, {Elvis}, \&
  {Loeb}}]{blecha13a}
{Blecha}, L., {Civano}, F., {Elvis}, M., \& {Loeb}, A. 2013, \mnras, 428, 1341.
  \eprint{1205.6202}

\bibitem[{{Blecha} et~al.(2011){Blecha}, {Cox}, {Loeb}, \&
  {Hernquist}}]{blecha11}
{Blecha}, L., {Cox}, T.~J., {Loeb}, A., \& {Hernquist}, L. 2011, \mnras, 412,
  2154. \eprint{1009.4940}

\bibitem[{{Blecha} \& {Loeb}(2008)}]{bleloe08}
{Blecha}, L., \& {Loeb}, A. 2008, \mnras, 390, 1311. \eprint{0805.1420}

\bibitem[{{Blecha} et~al.(2016){Blecha}, {Sijacki}, {Kelley}, {Torrey},
  {Vogelsberger}, {Nelson}, {Springel}, {Snyder}, \& {Hernquist}}]{blecha16}
{Blecha}, L., et.~al.
%{Blecha}, L., {Sijacki}, D., {Kelley}, L.~Z., {Torrey}, P., {Vogelsberger}, M.,
%  {Nelson}, D., {Springel}, V., {Snyder}, G., \& {Hernquist}, L. 
  2016, \mnras,
  456, 961. \eprint{1508.01524}

\bibitem[{{Blecha} et~al.(2018){Blecha}, {Snyder}, {Satyapal}, \&
  {Ellison}}]{blecha18}
{Blecha}, L., {Snyder}, G.~F., {Satyapal}, S., \& {Ellison}, S.~L. 2018,
  \mnras, 478, 3056. \eprint{1711.02094}

\bibitem[{{Bode} et~al.(2010){Bode}, {Haas}, {Bogdanovi{\'c}}, {Laguna}, \&
  {Shoemaker}}]{bode10}
{Bode}, T., et.~al.
%{Bode}, T., {Haas}, R., {Bogdanovi{\'c}}, T., {Laguna}, P., \& {Shoemaker}, D.
  2010, \apj, 715, 1117. \eprint{0912.0087}

\bibitem[{{Bonetti} et~al.(2016){Bonetti}, {Haardt}, {Sesana}, \&
  {Barausse}}]{bonett16}
{Bonetti}, M., {Haardt}, F., {Sesana}, A., \& {Barausse}, E. 2016, \mnras, 461,
  4419. \eprint{1604.08770}

\bibitem[{{Bonetti} et~al.(2018){Bonetti}, {Haardt}, {Sesana}, \&
  {Barausse}}]{bonett18}
--- 2018, \mnras, 477, 3910. \eprint{1709.06088}

\bibitem[{{Bowen} et~al.(2017){Bowen}, {Campanelli}, {Krolik}, {Mewes}, \&
  {Noble}}]{bowen17}
{Bowen}, D.~B., et.~al.
%{Bowen}, D.~B., {Campanelli}, M., {Krolik}, J.~H., {Mewes}, V., \& {Noble},
%  S.~C. 
  2017, \apj, 838, 42. \eprint{1612.02373}

\bibitem[{{Campanelli} et~al.(2007){Campanelli}, {Lousto}, {Zlochower}, \&
  {Merritt}}]{campan07b}
{Campanelli}, M., et.~al.
%{Campanelli}, M.,{Lousto}, C.~O., {Zlochower}, Y., \& {Merritt}, D. 
2007,
  Physical Review Letters, 98, 231102. \eprint{arXiv:gr-qc/0702133}

\bibitem[{{Chiaberge} et~al.(2017){Chiaberge}, {Ely}, {Meyer},
  {Georganopoulos}, {Marinucci}, {Bianchi}, {Tremblay}, {Hilbert}, {Kotyla},
  {Capetti}, {Baum}, {Macchetto}, {Miley}, {O'Dea}, {Perlman}, {Sparks}, \&
  {Norman}}]{chiabe17}
{Chiaberge}, M., et.~al.
%{Chiaberge}, M., {Ely}, J.~C., {Meyer}, E.~T., {Georganopoulos}, M.,
%  {Marinucci}, A., {Bianchi}, S., {Tremblay}, G.~R., {Hilbert}, B., {Kotyla},
%  J.~P., {Capetti}, A., {Baum}, S.~A., {Macchetto}, F.~D., {Miley}, G.,
%  {O'Dea}, C.~P., {Perlman}, E.~S., {Sparks}, W.~B., \& {Norman}, C. 
  2017,
  \aap, 600, A57. \eprint{1611.05501}

\bibitem[{{Civano} et~al.(2012){Civano}, {Elvis}, {Lanzuisi}, {Aldcroft},
  {Trichas}, {Bongiorno}, {Brusa}, {Blecha}, {Comastri}, {Loeb}, {Salvato},
  {Fruscione}, {Koekemoer}, {Komossa}, {Gilli}, {Mainieri}, {Piconcelli}, \&
  {Vignali}}]{civano12}
{Civano}, F., et.~al.
%{Civano}, F., {Elvis}, M., {Lanzuisi}, G., {Aldcroft}, T., {Trichas}, M.,
%  {Bongiorno}, A., {Brusa}, M., {Blecha}, L., {Comastri}, A., {Loeb}, A.,
%  {Salvato}, M., {Fruscione}, A., {Koekemoer}, A., {Komossa}, S., {Gilli}, R.,
%  {Mainieri}, V., {Piconcelli}, E., \& {Vignali}, C. 
2012, \apj, 752, 49.
  \eprint{1205.0815}

\bibitem[{{Civano} et~al.(2010){Civano}, {Elvis}, {Lanzuisi}, {Jahnke},
  {Zamorani}, {Blecha}, {Bongiorno}, {Brusa}, {Comastri}, {Hao}, {Leauthaud},
  {Loeb}, {Mainieri}, {Piconcelli}, {Salvato}, {Scoville}, {Trump}, {Vignali},
  {Aldcroft}, {Bolzonella}, {Bressert}, {Finoguenov}, {Fruscione}, {Koekemoer},
  {Cappelluti}, {Fiore}, {Giodini}, {Gilli}, {Impey}, {Lilly}, {Lusso},
  {Puccetti}, {Silverman}, {Aussel}, {Capak}, {Frayer}, {Le Floch},
  {McCracken}, {Sanders}, {Schiminovich}, \& {Taniguchi}}]{civano10}
{Civano}, F., et.~al.
%{Civano}, F., {Elvis}, M., {Lanzuisi}, G., {Jahnke}, K., {Zamorani}, G.,
%  {Blecha}, L., {Bongiorno}, A., {Brusa}, M., {Comastri}, A., {Hao}, H.,
%  {Leauthaud}, A., {Loeb}, A., {Mainieri}, V., {Piconcelli}, E., {Salvato}, M.,
%  {Scoville}, N., {Trump}, J., {Vignali}, C., {Aldcroft}, T., {Bolzonella}, M.,
%  {Bressert}, E., {Finoguenov}, A., {Fruscione}, A., {Koekemoer}, A.~M.,
%  {Cappelluti}, N., {Fiore}, F., {Giodini}, S., {Gilli}, R., {Impey}, C.~D.,
%  {Lilly}, S.~J., {Lusso}, E., {Puccetti}, S., {Silverman}, J.~D., {Aussel},
%  H., {Capak}, P., {Frayer}, D., {Le Floch}, E., {McCracken}, H.~J., {Sanders},
%  D.~B., {Schiminovich}, D., \& {Taniguchi}, Y. 
  2010, \apj, 717, 209.
  \eprint{1003.0020}

\bibitem[{{Condon} et~al.(2017){Condon}, {Darling}, {Kovalev}, \&
  {Petrov}}]{condon17}
{Condon}, J.~J., {Darling}, J., {Kovalev}, Y.~Y., \& {Petrov}, L. 2017, \apj,
  834, 184. \eprint{1606.04067}

\bibitem[{{Decarli} et~al.(2014){Decarli}, {Dotti}, {Mazzucchelli}, {Montuori},
  \& {Volonteri}}]{decarl14}
{Decarli}, R., et.~al.
%{Decarli}, R., {Dotti}, M., {Mazzucchelli}, C., {Montuori}, C., \& {Volonteri},
%  M. 
  2014, \mnras, 445, 1558. \eprint{1409.1585}

\bibitem[{{Eracleous} et~al.(2012){Eracleous}, {Boroson}, {Halpern}, \&
  {Liu}}]{eracle12}
{Eracleous}, M., {Boroson}, T.~A., {Halpern}, J.~P., \& {Liu}, J. 2012, \apjs,
  201, 23

\bibitem[{{Farris} et~al.(2014){Farris}, {Duffell}, {MacFadyen}, \&
  {Haiman}}]{farris14}
{Farris}, B.~D., {Duffell}, P., {MacFadyen}, A.~I., \& {Haiman}, Z. 2014, \apj,
  783, 134. \eprint{1310.0492}

\bibitem[{{Gualandris} \& {Merritt}(2008)}]{guamer08}
{Gualandris}, A., \& {Merritt}, D. 2008, \apj, 678, 780.
  \eprint{arXiv:0708.0771}

\bibitem[{{Hoffman} \& {Loeb}(2007)}]{hofloe07}
{Hoffman}, L., \& {Loeb}, A. 2007, \mnras, 377, 957.
  \eprint{arXiv:astro-ph/0612517}

\bibitem[{{Jonker} et~al.(2010){Jonker}, {Torres}, {Fabian}, {Heida},
  {Miniutti}, \& {Pooley}}]{jonker10}
{Jonker}, P.~G., et.~al.
%{Jonker}, P.~G., {Torres}, M.~A.~P., {Fabian}, A.~C., {Heida}, M., {Miniutti},
%  G., \& {Pooley}, D. 
  2010, \mnras, 407, 645. \eprint{1004.5379}

\bibitem[{{Kalfountzou} et~al.(2017){Kalfountzou}, {Santos Lleo}, \&
  {Trichas}}]{kalfou17}
{Kalfountzou}, E., {Santos Lleo}, M., \& {Trichas}, M. 2017, \apjl, 851, L15.
  \eprint{1712.03909}

\bibitem[{{Kelley} et~al.(2017){Kelley}, {Blecha}, \& {Hernquist}}]{kelley17a}
{Kelley}, L.~Z., {Blecha}, L., \& {Hernquist}, L. 2017, \mnras, 464, 3131.
  \eprint{1606.01900}


\bibitem[{{Kocevski} et~al.(2015){Kocevski}, {Brightman}, {Nandra},
  {Koekemoer}, {Salvato}, {Aird}, {Bell}, {Hsu}, {Kartaltepe}, {Koo}, {Lotz},
  {McIntosh}, {Mozena}, {Rosario}, \& {Trump}}]{kocevs15}
{Kocevski}, D.~D., et.~al.
%{Kocevski}, D.~D., {Brightman}, M., {Nandra}, K., {Koekemoer}, A.~M.,
%  {Salvato}, M., {Aird}, J., {Bell}, E.~F., {Hsu}, L.-T., {Kartaltepe}, J.~S.,
%  {Koo}, D.~C., {Lotz}, J.~M., {McIntosh}, D.~H., {Mozena}, M., {Rosario}, D.,
%  \& {Trump}, J.~R. 
  2015, \apj, 814, 104. \eprint{1509.03629}

\bibitem[{{Komossa} et~al.(2008){Komossa}, {Zhou}, \& {Lu}}]{komoss08}
{Komossa}, S., {Zhou}, H., \& {Lu}, H. 2008, \apjl, 678, L81.
  \eprint{arXiv:0804.4585}

\bibitem[{{Koss} et~al.(2014){Koss}, {Blecha}, {Mushotzky}, {Hung}, {Veilleux},
  {Trakhtenbrot}, {Schawinski}, {Stern}, {Smith}, {Li}, {Man}, {Filippenko},
  {Mauerhan}, {Stanek}, \& {Sanders}}]{koss14}
{Koss}, M., et.~al.
%{Koss}, M., {Blecha}, L., {Mushotzky}, R., {Hung}, C.~L., {Veilleux}, S.,
%  {Trakhtenbrot}, B., {Schawinski}, K., {Stern}, D., {Smith}, N., {Li}, Y.,
%  {Man}, A., {Filippenko}, A.~V., {Mauerhan}, J.~C., {Stanek}, K., \&
%  {Sanders}, D. 
  2014, \mnras, 445, 515. \eprint{1401.6798}

\bibitem[{{Kulkarni} \& {Loeb}(2012)}]{kulloe12}
{Kulkarni}, G., \& {Loeb}, A. 2012, \mnras, 422, 1306. \eprint{1107.0517}

\bibitem[{{Kumar} et~al.(2015){Kumar}, {Gezari}, {Heinis}, {Chornock},
  {Berger}, {Rest}, {Huber}, {Foley}, {Narayan}, {Marion}, {Scolnic},
  {Soderberg}, {Lawrence}, {Stubbs}, {Kirshner}, {Riess}, {Smartt}, {Smith},
  {Wood-Vasey}, {Burgett}, {Chambers}, {Flewelling}, {Kaiser}, {Metcalfe},
  {Price}, {Tonry}, \& {Wainscoat}}]{kumar15}
%{Kumar}, S., {Gezari}, S., {Heinis}, S., {Chornock}, R., {Berger}, E., {Rest},
%  A., {Huber}, M.~E., {Foley}, R.~J., {Narayan}, G., {Marion}, G.~H.,
%  {Scolnic}, D., {Soderberg}, A., {Lawrence}, A., {Stubbs}, C.~W., {Kirshner},
%  R.~P., {Riess}, A.~G., {Smartt}, S.~J., {Smith}, K., {Wood-Vasey}, W.~M.,
%  {Burgett}, W.~S., {Chambers}, K.~C., {Flewelling}, H., {Kaiser}, N.,
%  {Metcalfe}, N., {Price}, P.~A., {Tonry}, J.~L., \& {Wainscoat}, R.~J. 
 {Kumar}, S., et.~al. 2015,
  \apj, 802, 27. \eprint{1501.01314}


\bibitem[{{Lena} et~al.(2014){Lena}, {Robinson}, {Marconi}, {Axon}, {Capetti},
  {Merritt}, \& {Batcheldor}}]{lena14}
{Lena}, D., et.~al.
%{Lena}, D., {Robinson}, A., {Marconi}, A., {Axon}, D.~J., {Capetti}, A.,
%  {Merritt}, D., \& {Batcheldor}, D. 
  2014, \apj, 795, 146. \eprint{1409.3976}

\bibitem[{{Loeb}(2007)}]{loeb07}
{Loeb}, A. 2007, Physical Review Letters, 99, 041103.
  \eprint{arXiv:astro-ph/0703722}


\bibitem[{{L{\'o}pez-Navas} \& {Prieto}(2018)}]{loppri18}
{L{\'o}pez-Navas}, E., \& {Prieto}, M.~A. 2018, \mnras, 480, 4099. \eprint{1808.04123}

\bibitem[{{Lousto} et~al.(2010){Lousto}, {Campanelli}, {Zlochower}, \&
  {Nakano}}]{lousto10}
{Lousto}, C.~O., et.~al.
%{Lousto}, C.~O., {Campanelli}, M., {Zlochower}, Y., \& {Nakano}, H. 
2010,
  Classical and Quantum Gravity, 27, 114006. \eprint{0904.3541}

\bibitem[{{Madau} \& {Quataert}(2004)}]{madqua04}
{Madau}, P., \& {Quataert}, E. 2004, \apjl, 606, L17.
  \eprint{arXiv:astro-ph/0403295}

\bibitem[{{Markakis} et~al.(2015){Markakis}, {Dierkes}, {Eckart}, {Nishiyama},
  {Britzen}, {Garc{\'{\i}}a-Mar{\'{\i}}n}, {Horrobin}, {Muxlow}, \&
  {Zensus}}]{markak15}
{Markakis}, K., et.~al.
%{Markakis}, K., {Dierkes}, J., {Eckart}, A., {Nishiyama}, S., {Britzen}, S.,
%  {Garc{\'{\i}}a-Mar{\'{\i}}n}, M., {Horrobin}, M., {Muxlow}, T., \& {Zensus},
%  J.~A. 
  2015, \aap, 580, A11. \eprint{1504.03691}

\bibitem[{{Novak} et~al.(2015){Novak}, {Smol{\v c}i{\'c}}, {Civano}, {Bondi},
  {Ciliegi}, {Wang}, {Loeb}, {Banfield}, {Bourke}, {Elvis}, {Hallinan},
  {Intema}, {Kl{\"o}ckner}, {Mooley}, \& {Navarrete}}]{novak15}
{Novak}, M., et.~al.
%{Novak}, M., {Smol{\v c}i{\'c}}, V., {Civano}, F., {Bondi}, M., {Ciliegi}, P.,
%  {Wang}, X., {Loeb}, A., {Banfield}, J., {Bourke}, S., {Elvis}, M.,
%  {Hallinan}, G., {Intema}, H.~T., {Kl{\"o}ckner}, H.-R., {Mooley}, K., \&
%  {Navarrete}, F. 
  2015, \mnras, 447, 1282. \eprint{1412.0004}

\bibitem[{{Ricci} et~al.(2017){Ricci}, {Bauer}, {Treister}, {Schawinski},
  {Privon}, {Blecha}, {Arevalo}, {Armus}, {Harrison}, {Ho}, {Iwasawa},
  {Sanders}, \& {Stern}}]{ricci17}
{Ricci}, C., et.~al.
%{Ricci}, C., {Bauer}, F.~E., {Treister}, E., {Schawinski}, K., {Privon}, G.~C.,
%  {Blecha}, L., {Arevalo}, P., {Armus}, L., {Harrison}, F., {Ho}, L.~C.,
%  {Iwasawa}, K., {Sanders}, D.~B., \& {Stern}, D. 
  2017, \mnras, 468, 1273.
  \eprint{1701.04825}

\bibitem[{{Robinson} et~al.(2010){Robinson}, {Young}, {Axon}, {Kharb}, \&
  {Smith}}]{robins10}
{Robinson}, A., et.~al.
%{Robinson}, A., {Young}, S., {Axon}, D.~J., {Kharb}, P., \& {Smith}, J.~E.
  2010, \apjl, 717, L122. \eprint{1006.0993}

\bibitem[{{Runnoe} et~al.(2017){Runnoe}, {Eracleous}, {Pennell}, {Mathes},
  {Boroson}, {Sigursson}, {Bogdanovi{\'c}}, {Halpern}, {Liu}, \&
  {Brown}}]{runnoe17}
{Runnoe}, J.~C., et.~al.
%{Runnoe}, J.~C., {Eracleous}, M., {Pennell}, A., {Mathes}, G., {Boroson}, T.,
%  {Sigursson}, S., {Bogdanovi{\'c}}, T., {Halpern}, J.~P., {Liu}, J., \&
%  {Brown}, S. 
  2017, \mnras, 468, 1683. \eprint{1702.05465}

\bibitem[{{Seth} et~al.(2014){Seth}, {van den Bosch}, {Mieske}, {Baumgardt},
  {Brok}, {Strader}, {Neumayer}, {Chilingarian}, {Hilker}, {McDermid},
  {Spitler}, {Brodie}, {Frank}, \& {Walsh}}]{seth14}
{Seth}, A.~C., et.~al. 
%{Seth}, A.~C., {van den Bosch}, R., {Mieske}, S., {Baumgardt}, H., {Brok},
%  M.~D., {Strader}, J., {Neumayer}, N., {Chilingarian}, I., {Hilker}, M.,
%  {McDermid}, R., {Spitler}, L., {Brodie}, J., {Frank}, M.~J., \& {Walsh},
%  J.~L. 
  2014, \nat, 513, 398. \eprint{1409.4769}

\bibitem[{{Sijacki} et~al.(2011){Sijacki}, {Springel}, {Haehnelt}}]{sijack11}
{{Sijacki}, D. and {Springel}, V. and {Haehnelt}, M.~G.}
2011, \mnras, 414, 3656. \eprint{1008.3313}

\bibitem[{{Veilleux} et~al.(2009){Veilleux}, {Rupke}, {Kim}, {Genzel}, {Sturm},
  {Lutz}, {Contursi}, {Schweitzer}, {Tacconi}, {Netzer}, {Sternberg}, {Mihos},
  {Baker}, {Mazzarella}, {Lord}, {Sanders}, {Stockton}, {Joseph}, \&
  {Barnes}}]{veille09}
{Veilleux}, S., et.~al.
%{Veilleux}, S., {Rupke}, D.~S.~N., {Kim}, D.-C., {Genzel}, R., {Sturm}, E.,
%  {Lutz}, D., {Contursi}, A., {Schweitzer}, M., {Tacconi}, L.~J., {Netzer}, H.,
%  {Sternberg}, A., {Mihos}, J.~C., {Baker}, A.~J., {Mazzarella}, J.~M., {Lord},
%  S., {Sanders}, D.~B., {Stockton}, A., {Joseph}, R.~D., \& {Barnes}, J.~E.
  2009, \apjs, 182, 628-666. \eprint{0905.1577}

\bibitem[{{Volonteri}(2007)}]{volont07}
{Volonteri}, M. 2007, \apjl, 663, L5. \eprint{arXiv:astro-ph/0703180}

\bibitem[{{Volonteri} \& {Madau}(2008)}]{volmad08}
{Volonteri}, M., \& {Madau}, P. 2008, \apjl, 687, L57. \eprint{0809.4007}

\end{thebibliography}

% For non-BibTex:
%\begin{thebibliography}{}
%\bibitem[... (...)]{...} ...
%\end{thebibliography}

\end{document}